\documentclass{article}
\usepackage{graphicx}
\usepackage{hyperref}
\usepackage{authblk}
\usepackage{algorithm}
\usepackage{algorithmicx}
\usepackage{algpseudocode}
\usepackage{amsmath}
\numberwithin{equation}{section}
\usepackage{indentfirst}
\usepackage{amsfonts}
\usepackage{framed}
\begin{document}
\title{\Large Change of measure under the hard-to-borrow model}
\author[a]{Peng Liu \thanks{Corresponding author: mf1702010@smail.nju.edu.cn, peng.liu.john@gmail.com}}
\affil[a]{Department of Finance and Insurance, Nanjing University}

\maketitle
    \begin{abstract}
    As the Securities and Exchange Commission(SEC) has implemented a new regulation on short-sellings, short-sellers are required to repurchase stocks once the clearing risk rises to a certain level.
	Avellaneda and Lipkin proposed a fully coupled SDE system to describe the mechanism which is referred as Hard-To-Borrow(HTB) models.
	Guiyuan Ma obtained the PDE system for both American and European options.
	There is a technical error in Guiyuan Ma where two correlated Brownian motion should be converted before change of measure.
	In this paper, I will provide supplement conditions. 
    \end{abstract}
	
\section{Introduction}
	
    \par
    One of the most important tasks in mathematical finance is to determine an option's value.
    The foundation of modern option pricing theory was attributed to Black Scholes\cite{Black1973}, Merton\cite{Merton1973} who first proposed the analytical solution for European options.
    Black and Scholes\cite{Black1973} assumed the market is complete, meaning the short-sellings on underlying assets are permitted without costs.
    However, short-sellings are restricted in many developed markets, or even forbidden in most emerging ones. 
    Generally, short-selling requires borrowing stocks from others where different stakeholders are involved.
    The broker is going to arrange a buyer once the stock has been indicated by the short-seller.
    Then the short-seller need return stocks to the clearing firm within a negotiated period.
    Stocks can be borrowed from a stock-loan desk.
    The availability of stocks for borrowing depends on market conditions.
    There are many stocks which can be easily borrowed without any lending fees, which exactly fits one of Black-Scholes assumptions, while others could be in short supply.
    In the latter case, short-selling those stocks, referred as \emph{hard-to-borrow}(HTB) stocks, could be quite costly.
	
    \par
	Normally, there would not be any unexpected cash flows before the settlement.
    Regulators have paid attention on the clearing risk in short-sellings.
    Since the \emph{buy-in} mechanism, has been implemented by the Securities and Exchange Commission(SEC), 
    clearing firms representing short-sellers are required to repurchase stocks to cover shortfalls once the clearing risk reaches to a certain level.
    To describe the buy-in mechanism associated with hard-to-borrow stocks, Avellaneda and Lipkin\cite{Avellaneda2009} proposed a dynamic model (it is referred to as the HTB model) by introducing two coupled SDEs.
    The HTB model has attracted attention from different aspects.
	Guiyuan Ma and Song-ping Zhu \cite{Ma2018} studied the early exercise of American option on the HTB model.
	They confirmed that it is the lending fee that results in the early exercise of American call options and we shall also demonstrate to what extent lending fees have affected the early exercise decision.
	Yong Chen and Jingtang Ma\cite{Chen2018} introduced a new expansion approach to deal with the spatial delay term.
	They compared the modified Laplace transform method with the time-stepping finite difference method.
	The numerical comparison indicates that the modified Laplace transform method outperforms the modified FDMs.
	Guiyuan Ma, Song-ping Zhu, Wenting Chen\cite{Ma2019} studied European call option pricing problem under the HTB model.
	Through their numerical results, they find that the semi-explicit formula is a good approximate solution when the coupling parameter is small. 
	When the stock price and the buy-in rate are significantly coupled, the PDE approach is preferred to solve the option pricing problem under the full hard-to-borrow model.

    \par
    The contribution of this paper is amending a technical error during change of measure.
	Avellaneda and Lipkin \cite{Avellaneda2009} claim that the correlation of two Brownian motion is irrelevant.
    The situation where two Brownian motion $W(t),Z(t)$ are correlated has not been discussed which is contrary to the footprint of [].
    In the following part, I will provide a mathematical proof that the change of measure depends on the covariance between $W(t)$ and $Z(t)$.
    To change of measure under multi-dimension \emph{Girsanov} theorem, multi-dimension Brownian motion $B(t)=(B_{1}(t),B_{2}(t))$ are introduced to convert the correlated Brownian motions $W(t),Z(t)$ into a two-dimension Brownian motion.
	My future work will focus on fast simulation for the HTB model.
	
\section{The Coupled SDE system}
	
    \par
    Avellaneda and Lipkin\cite{Avellaneda2009} modelled the buy-in mechanism with two coupled SDEs.
    
    \begin{equation}
    \left\{
    \begin{array}{lr}
    \frac{dS_{t}}{S_{t}}=\sigma dW_{t}+\gamma \lambda_{t}dt-\gamma dN_{\lambda_{t}}(t) & \label{eq:1} \\
    dx_{t}=\kappa dZ_{t}+\alpha(\overline{x}-x_{t})dt+\beta\frac{dS_{t}}{S_{t}},x_{t}=ln(\dfrac{\lambda_{t}}{\lambda_{0}})&\\
    \end{array}
    \right.
    \end{equation}
    
    The first equation describes the logarithmized price process with a Brownian motion $W_{t}$ and a compensated \emph{Poisson} process $\gamma \lambda_{t}dt-\gamma dN_{\lambda_{t}}(t)$.
    Short-sellers would suffer from "squeezes" triggered by buy-ins.
    The profit or loss during a time interval \emph{(t,t+dt)} affected by buy-ins can be represented as
    \begin{equation}
    \begin{aligned} 
    &PNL=-dS_{t}-\xi \gamma S_{t}=-S_{t}(\sigma dW_{t}+\lambda_{t}\gamma dt) \\
    &Prob.\{\xi=0\}=1-\lambda_{t}dt+o(dt)\\
    &Prob.\{\xi=1\}=\lambda_{t}dt+o(dt)
    \end{aligned} 
    \end{equation}
   
    The second Mean-Reverting process characterizes the logarithmized \emph{Poisson} intensity $\lambda$ of the previous equation and another Brownian motion $Z_{t}$ is introduced.
    Two Brownian motions $W_{t},Z_{t}$ with volatility $\sigma,\kappa$ respectively are correlated with coefficient $\rho=cov(W_{t},Z_{t})$ while another parameter $\beta$ determines how deeply two equations are coupled.
    $\gamma$ is the price elasticity of demand due to buy-ins.
	$\alpha$ determines the reverting speed and $\overline{x}$ is the long-term level of $x$.
		
\section{Change Of Measure}

	\subsection{Arbitrage-free Measure}			
    \par
    The measure change of the HTB models was initially proposed by Avellaneda and Lipkin.	
    Avellaneda and Lipkin\cite{Avellaneda2009} assumed that short-sellers do not benefit from the downward jumps because short-sellers are no longer short by the time the buy-in
    is completed.
    Since jumps and buy-ins occur with frequency $\lambda_{t}$, the expected economic gain is $\lambda_{t}\gamma S_{t}$. 
	$\lambda_{t} \gamma$ can be viewed as the cost-of-carry for borrowing the stock. 
    Statistically the economic costs of paying rent or risking buy-ins are equivalent.
    In particular, the cost of carrying (or financing) stock can be quantified in terms of $\lambda$ and the interest rate.
    
    \par
    An arbitrage-free pricing measure associated with the physical process was introduced.
    After change of measure, the first equation of \ref{eq:1} should take the form:
	\begin{equation}
	\frac{dS_{t}}{S_{t}}=\sigma dW_{t}+rdt-\gamma dN_{\lambda_{t}}(t)  
	\end{equation}
	where $r$ is the instantaneous interest rate. 
	The absence of the drift term $\lambda_{t}\gamma$ in this last equation is due to the fact that, under an arbitrage pricing measure,
	the price process adjusted for dividends and interest is a martingale.

	\subsection{Risk-neutral Measure}	
	\par
	The risk-neutral measure was formally obtained by Guiyuan Ma\cite{Ma2019}.
	To conduct measure transform, Guiyuan Ma defined two new processes as
	
	\begin{align}
	\widetilde{W_{t}}&=W_{t}+\int_{0}^{t}\frac{\gamma\lambda_{l}-r}{\sigma}dl\\
	\widetilde{Z_{t}}&=Z_{t}+\int_{0}^{t}\frac{\alpha z(l,x_{l},S_{l})}{\kappa}dl
	\end{align}
	
	The risk-neutral measure was defined by the \emph{Randon-Nikodym} derivative 
	
	\begin{flalign}
	\frac{d\mathbb{Q}}{d\mathbb{P}} \Bigg|_{t}=exp\{-\int_{0}^{t}[\frac{\gamma\lambda_{l}-r}{\sigma}+\frac{\alpha z(l,x_{l},S_{l})}{\kappa}]dW_{l}- \frac{1}{2}\int_{0}^{t}[\frac{(\gamma\lambda_{l}-r)^{2}}{\sigma^{2}}+\frac{\alpha^{2} z^{2}(l,x_{l},S_{l})}{\kappa^{2}}]dl\}
	\end{flalign}
	
	\par
	Guiyuan Ma explained that $z(t,x,S)$ is an arbitrary function for the uncertainty raised by buy-ins.
	Any source of uncertainty needs to be compensated by the associated market price of risk or risk premium. 	
	In the classic Black–Scholes model\cite{Black1973},\cite{Merton1973}, the market price of risk for the underlying is $\frac{\mu-r}{\sigma}$.
	In the Heston model\cite{Heston1993}, an additional source uncertainty is introduced by the stochastic volatility and an additional market price of volatility risk is defined through an arbitrary function, i.e., $\lambda(t, S, v)$
	In the HTB model, the new buy-in process also brings in an additional source of uncertainty and the corresponding market price of buy-in risk is represented by the function $z(t,x,S)$.	
	HTB model operates in an incomplete market where it is impossible to perfectly hedge a portfolio composed of hard-to-borrow stocks and a unique risk-neutral measure does not exist.
	The market price of buy-in risk in the HTB model should be determined by market data.
	
\section{Two Correlated Brownian Motion}     
    
    \par
    Avellaneda and Lipkin\cite{Avellaneda2009} claimed that the correlation of two Brownian motion $W_{t},Z_{t}$ in \ref{eq:1} is irrelevant.
    The mathematical result obtained by Guiyuan Ma\cite{Ma2019} also neglect the situation where two Brownian motion are correlated.
    The contribution of my work is to give supplement conditions before change of measure.  
	
    \par 
    For convenience, we denote $\Gamma(l)=\frac{\gamma\lambda_{l}-r}{\sigma}$,$\Theta(l)=\frac{\alpha z(l,x_{l},S_{l})}{\kappa}$

    \par 
    To change of measure under multi-dimension Girsanov theorem, multi-dimension Brownian motion $B(t)=(B_{1}(t),B_{2}(t))$ are introduced to convert the correlated Brownian motions $W(t),Z(t)$ into a two-dimension Brownian motion.

    \par 
    \textbf{\emph{Proposition 1}}
    \emph{$B_{1}(t),B_{2}(t)$ are two independent Brownian motion,$Y_{1}(t)=B_{1}(t)$ $X_{1}(t),X_{2}(t)$ are two correlated Brownian motion with covariance $\rho$}

    \par
    \textbf{\emph{Proof}}
    \begin{equation}
    \begin{aligned}
    X_{1}(t)&=B_{1}(t)\\
    X_{2}(t)&=\rho B_{1}(t)+\sqrt{1-\rho^{2}}B_{2}(t)
    \end{aligned}
    \end{equation}
    
    \begin{equation}
    \begin{aligned}
    dX_{2}(t)dX_{2}(t)&=\rho^{2}dB_{1}(t)dB_{1}(t)+2\rho\sqrt{1-\rho^{2}}dB_{1}(t)dB_{2}(t)\\
                &+(1-\rho^{2})dB_{2}(t)dB_{2}(t)\\
                &=\rho^{2}dt+(1-\rho^{2})dt=dt
    \end{aligned}
    \end{equation}

    \begin{equation}
    \begin{aligned}
    dX_{1}(t)dX_{2}(t)&=\rho^{2}dB_{1}(t)dB_{1}(t)+\sqrt{1-\rho^{2}}dB_{1}(t)dB_{2}(t)\\
                      &=\rho dt
    \end{aligned}
    \end{equation}

    \textbf{\emph{Girsanov theorem}}\cite{Oksendal} \emph{Let} $Y(t)\in\mathbb{R}^{n}$ \emph{be an} $It\hat{o}\ process$ \emph{of the form}
    
    \begin{equation}
    dY(t)=\beta(t,\omega)dt+\theta(t,\omega)dB(t)\nonumber
    \end{equation}
    
    \emph{where} $B(t)\in\mathbb{R}^{m}$,$\beta(t,\omega)\in\mathbb{R}^{n}$ and $\theta(t,\omega)\in\mathbb{R}^{n\times m}$.\emph{Suppose there exist processes} $u(t,\omega)\in\mathcal{W}_{\mathcal{H}}^{m}$ \emph{and} $\alpha(t,\omega)\in\mathcal{W}_{\mathcal{H}}^{n}$ \emph{such that}
    
    \begin{equation}
    \theta(t,\omega)u(t,\omega)=\beta(t,\omega)-\alpha(t,\omega)\nonumber
    \end{equation}
    
    \emph{Put}\\
    
    \begin{equation}
    M_{t}=exp\{-\int_{0}^{t}u(s,\omega)dB_{s} -\frac{1}{2}\int_{0}^{t}u^{2}(s,\omega)ds\};\quad t\leq T
    \end{equation}
    
    \emph{and}\\
    
    \begin{equation}
    dQ(\omega)=M_{T}(\omega)dP(\omega)\qquad on \quad \mathcal{F}_{T}^{(m)}
    \end{equation}
    
    \emph{Assume that} $M_{t}$ \emph{is a martingale($\omega.r.t.\ \mathcal{F}_{t}^{(n)}$and P)}. \emph{Then Q is a probability measure on $\mathcal{F}_{T}^{(m)}$,the process}\\
    
    \begin{equation}
    \hat{B}(t):=\int_{0}^{t}u(s,\omega)ds+B(t);\qquad t\leq T
    \end{equation}
    
    \emph{is a Brownian motion $\omega.r.t$.Q and in terms of $\hat{B}(t)$ the process Y(t) has the stochastic integral representation}\\
    
    \begin{equation}
    dY(t)=\alpha(t,\omega)dt+\theta(t,\omega)d\hat{B}(t)
    \end{equation}
    
    \emph{If n=m and $\theta \in\mathbb{R}^{n\times n} $ is invertible, then the process $u(t,\omega)$ satisfying is given by}
    
    \begin{equation}
    u(t,\omega)=\theta^{-1}(t,\omega)[\beta(t,\omega)-\alpha(t,\omega)]
    \end{equation}

    \par
    \textbf{\emph{Proposition 2}}
    \emph{The stochastic process $Y(t)=(\widetilde{W_{t}},\widetilde{Z_{t}})\quad(-1< \rho< 1,0<t\leq T)$ is a local martingale under the measure $\mathbb{Q}$ defined by}\\
    
    \begin{equation}
    \frac{d\mathbb{Q}}{d\mathbb{P}}\Bigg|_{t}=exp\{-\int_{0}^{t}\Gamma(u)dB_{1}(u)-\int_{0}^{t}\frac{\Theta(u)-\rho\Gamma(u)}{\sqrt{1-\rho^{2}}}dB_{2}(u)
    -\frac{1}{2}\int_{0}^{t}\frac{\Theta^{2}(u)+\Gamma^{2}(u)-2\rho\Gamma(u)\Theta(u)}{1-\rho^2}du\}
    \end{equation}
    
    \par 
    \textbf{\emph{Proof}}\\

    \begin{equation}
    \begin{aligned}
    Y_{1}(t)&=B_{1}(t)+\int_{0}^{t}\Gamma(u)du\\
    Y_{2}(t)&=\rho B_{1}(t)+\sqrt{1-\rho^{2}}B_{2}(t)+\int_{0}^{t}\Theta(u)du
    \end{aligned}
    \end{equation}

    \begin{equation}
    dY(t) = \left[ \begin{array}{c}
    {\Gamma(t)}\\
    {\Theta(t)}\\
    \end{array} \right]dt + \left[ \begin{array}{ll}
    1 & \qquad 0 \\
    \rho & \sqrt{1 - \rho^{2}} \\
    \end{array} \right]\left[ \begin{array}{c}
    {dB_{1}(t)} \\
    {dB_{2}(t)} \\
    \end{array} \right]
    \end{equation}
    
	In order to get risk-neutral measure, the process $\alpha(t,\omega)$ is chosen to be zero. Choose $\alpha(t,\omega)=0$. Then equation gets the form:
    
    \begin{equation}
    \left[ \begin{array}{ll}
    1 & \qquad 0 \\
    \rho & \sqrt{1 - \rho^{2}} \\
    \end{array} \right]
    \left[ \begin{array}{l}
    {u_{1}(t)} \\
    {u_{2}(t)} \\
    \end{array} \right]=
    \left[\begin{array}{l}
    {\Gamma(t)} \\
    {\Theta(t)} \\
    \end{array} \right]
    \end{equation}
	which has the unique solution
	
	\begin{equation}
	u_{1}(t)=\Gamma(t),u_{2}(t)=\frac{-\rho\Gamma(t)+\Theta(t)}{\sqrt{1-\rho^{2}}}
	\end{equation}
	
	Then the risk-neutral measure defined by \emph{Randon-Nikodym} derivative is
    
    \begin{equation}
    \begin{aligned}
    \frac{d\mathbb{Q}}{d\mathbb{P}}\Bigg|_{t}&=exp\{-\int_{0}^{t}u(s,\omega)dB_{s} -\frac{1}{2}\int_{0}^{t}u^{2}(s,\omega)ds\}\\
    &=exp\{-\int_{0}^{t}\Gamma(u)dB_{1}(u)-\int_{0}^{t}\frac{\Theta(u)-\rho\Gamma(u)}{\sqrt{1-\rho^{2}}}dB_{2}(u)
    -\frac{1}{2}\int_{0}^{t}\frac{\Theta^{2}(u)+\Gamma^{2}(u)-2\rho\Gamma(u)\Theta(u)}{1-\rho^2}du\}
    \end{aligned}
    \end{equation}

\section{Conclusion}
	\par
	In this paper, I provide the missing conditions before change of measure in the HTB model.
	The correlation between of two Brownian motion requires further work.
	My future work will extend the Mote-Carlo simulation of the HTB model.
\clearpage
\bibliographystyle{unsrt}
\bibliography{Collection1}

\end{document}